\documentstyle[12pt]{article}
\def\be{\begin{equation}}
\def\ee{\end{equation}}
\def\bea{\begin{eqnarray}}
\def\eea{\end{eqnarray}}

\begin{document}

\begin{center}
{\Large{\bf Rotating and Moving D-Branes in the\\
Presence of Various Background Fields}}

\vskip .5cm
{\large Farzin Safarzadeh-Maleki and Davoud Kamani}
\vskip .1cm
{\it Faculty of Physics, Amirkabir University of Technology
(Tehran Polytechnic)\\
P.O.Box: 15875-4413, Tehran, Iran}\\
{\sl e-mails: kamani@aut.ac.ir , f.safarzadeh@aut.ac.ir}\\
\end{center}

\begin{abstract}

We calculate the bosonic boundary state associated with a rotating
and moving D$p$-brane in the presence of the antisymmetric 
tensor field, a $U(1)$ gauge field and a tachyon field.
Rotation and motion are in the brane volume.
We reconstruct this boundary state via the group $PSL(2,R)$
to be applicable when the tachyon field is presented.
This modified boundary state enables us to calculate the
interaction amplitude between two parallel D$p$-branes with
rotation and motion. The long-range force of this interaction
will be obtained. The boundary state also enables us to 
investigate the tachyon condensation on a rotating and moving 
D$p$-brane.

\end{abstract}



\vskip .5cm

\newpage

\section{INTRODUCTION}

D-branes, as essential objects in string theory \cite{1,2},
have important applications in different aspects such as:
braneworld cosmology, stability of time dependent phenomena,
solving many timelike singularities of general relativity,
describing nonperturbative phenomena in string theory, higher
dimensional theories and so on. In addition, D-branes
with nonzero background internal fields have shown several
interesting properties \cite{3}-\cite{8}. For example, internal 
fields control the interactions of the branes and/or the background
tachyon field specifies instability of the branes.

On the other hand, we have the boundary state formalism
for describing the D-branes \cite{9}-\cite{16}.
This formalism is not only a useful tool in many
complicated situations, even when a clear spacetime is not
available, but also it can find the conformal duality which
is an important problem in revealing the underlying symmetry
of string theory. The overlap of two boundary states,
corresponding to two D-branes, through the closed string
propagator gives the interaction amplitude of the branes.
So far this adequate method has been applied to the stationary mixed
branes, moving branes with constant velocities and rotated
branes by an angle in the presence of different background
fields \cite{16}-\cite{20}. For the branes with the tachyon 
field see Refs. \cite{21}-\cite{31}.

The goal of this paper is to determine a general case including
rotating and moving D-branes in the presence of the following
background fields: the Kalb-Ramond field, $U(1)$ gauge fields
which live in the D-branes worldvolumes and tachyon fields.
We shall consider rotation of a brane in its volume, and its
velocity is along the brane directions. Because of the various
fields inside the branes there are preferred
directions and hence these rotations and motions are meaningful.
This generalized setup strongly affects interaction of the
branes.

In fact, the presence of the tachyon field, especially a tachyon
field with the quadratic function of the spacetime coordinates,
induces an off-shell theory. In this article we shall
use the prescription of Ref. \cite{27} to reconstruct
the boundary state associated with a D$p$-brane, in the presence
of the quadratic tachyon field. The modified boundary state will be
applied for computing the interaction amplitude of two D$p$-branes,
which exchange on-shell and off-shell closed strings. 

In the last few years significant steps have been made to
understand the open string tachyon dynamics and tachyon
condensation. These concepts have been studied by the first
quantized string theory \cite{32}-\cite{34}, by the cubic open string
field theory \cite{35}, by the RG flow method \cite{36}-\cite{38} and more
recently by the boundary string field theory
\cite{39}-\cite{41}. In these studies it has been shown that open
string tachyon condensation describes the decay of unstable
D-branes into the closed string vacuum (or to stable ones). In
this paper, we shall examine the tachyon condensation for a rotating-moving 
D$p$-brane. We observe that tachyon condensation can always make such 
branes to be unstable and hence reduces the brane dimension. 

This paper is organized as follows. In Sec. 2, the boundary
state associated with a rotating and moving D$p$-brane with
background fields will be constructed. In Sec. 3, the interaction of
two D$p$-branes will be studied. In Sec. 4, the instability of a
rotating-moving D$p$-brane due to the tachyon condensation
phenomena will be examined. Section 5 is devoted to the conclusions.
\section{THE BOUNDARY STATE}

\subsection{Extracting the boundary state from the action}

Our starting point is to determine the boundary action corresponding
to a rotating and moving D$p$-brane.
Therefore, we use the following sigma-model action for the 
closed string:
\bea
S =&-&\frac{1}{4\pi\alpha'} {\int}_\Sigma
d^{2}\sigma(\sqrt{-g}g^{ab}G_{\mu\nu}\partial_a X^{\mu}\partial_b
X^{\nu}+\varepsilon^{ab} B_{\mu\nu}\partial_a X^{\mu}\partial_b
X^{\nu})
\nonumber\\
&+&\frac{1}{2\pi\alpha'} {\int}_{\partial\Sigma}
d\sigma ( A_\alpha
\partial_{\sigma}X^{\alpha}+ \omega_{\alpha\beta}J^{\alpha\beta}_{\tau }
+T(X^\alpha)),
\eea
where $\Sigma$ is the closed string worldsheet
and $\partial\Sigma$  is its boundary. We shall
use $\{X^\alpha|\alpha =0, 1, \cdot \cdot \cdot ,p \}$ for
the worldvolume directions of the D$p$-brane and
$\{X^i| i= p+1, \cdot \cdot \cdot ,d-1\}$ for directions
perpendicular to it.
This action contains the Kalb-Ramond field $B_{\mu\nu}$,
a $U(1)$ gauge field $A_\alpha$ which lives in the worldvolume
of the brane, an $\omega$-term associated with the rotation
and motion of the brane and a tachyonic field. Note that the
tachyon and gauge potential are in the open string spectrum,
which are attached to the brane.

For simplifying the calculations the background fields $G_{\mu \nu}$
and $B_{\mu \nu}$ are considered to be constant. In addition, for
the $U(1)$ gauge field we apply the gauge
$A_{\alpha}=-\frac{1}{2}F_{\alpha \beta }X^{\beta}$ with
constant field strength. Besides, we use the following tachyon
profile $T=T_0 +\frac{1}{2}U_{\alpha\beta}X^{\alpha}X^{\beta}$,
where $T_0$ and the symmetric matrix $U_{\alpha\beta}$ are constant.
Finally, the $\omega$-term, which is responsible for the 
rotation and motion of the brane, contains the antisymmetric 
angular velocity
${\omega }_{\alpha \beta}$ and angular momentum density
$J^{\alpha \beta }_{\tau}$. Its explicit form is given by
${\omega }_{\alpha \beta}J^{\alpha
\beta }_{\tau }=2{\omega }_{\alpha \beta }X^{\alpha }{\partial }_{\tau
}X^{\beta }$. In fact, 
$\omega_{0 {\bar \alpha}}|_{{\bar \alpha} \neq 0}$ denotes the
velocity component 
of the brane along the direction $X^{\bar \alpha}$ while
$\omega_{{\bar \alpha}{\bar \beta}}$ represents its rotation.
Note that in the presence of the Kalb-Ramond field and the 
local gauge potential there are some preferred alignments in the brane.
This implies that the rotation and motion of the brane in its
volume are sensible.

Vanishing of the variation of the action with respect to
$X^{\mu }(\sigma ,\tau )$ gives the equation of motion and the
following boundary state equations
\bea
&~& [{(\eta }_{\alpha \beta }
+4{\omega }_{\alpha \beta }){\partial }_{\tau }X^{\beta }
+{{{\mathcal F}}_{{\mathbf \alpha \beta}}}{\partial }_{\sigma }X^{\beta }
+U_{\alpha \beta }X^{\beta }]|_{\tau =0}\ \ |B\rangle\ =0,
\nonumber\\
&~& {\delta X}^i|_{\tau =0}|B\rangle\ =0,
\eea
where the total field strength is
${\cal{F}}_{\alpha \beta}=F_{\alpha \beta}-B_{\alpha \beta}$.
For simplicity, we assumed that the following mixed elements
are zero $B_{\alpha i}=U_{\alpha i}=0$.

Now by using the closed string mode expansion
\bea
X^{\mu }(\sigma ,\tau )=x^{\mu }+l^2p^{\mu }\tau
+\frac{1}{2}il\sum_{m\ne 0}{\frac{1}{m}\ (\ {\alpha }^{\mu }_m
e^{-2im\left(\tau -\sigma \right)}
+}{\widetilde{\alpha }}^{\mu }_me^{-2im(\tau +\sigma )}),
\eea
the boundary state equations can be written in terms of the
oscillators
\bea
[{(\eta }_{\alpha \beta }+4{\omega }_{\alpha \beta })l^2p^{\beta }
+U_{\alpha \beta }x^{\beta }]{|B\rangle}^{\left(0\right)}\ =0,
\eea
\bea
(x^{i}-y^{i}){|B\rangle}^{\left(0\right)}\ =0,
\eea
\bea
[({\eta }_{\alpha \beta }+4{\omega }_{\alpha \beta }-
{{\mathcal F}}_{{\mathbf \alpha }{\mathbf \beta }}
+\frac{i}{2m}U_{\alpha \beta }){\alpha }^{\beta }_m
+{({\eta }_{\alpha \beta }+4{\omega }_{\alpha \beta }
+{{\mathcal F}}_{{\mathbf \alpha }
{\mathbf \beta }}\ -\frac{i}{2m}U_{\alpha \beta })}
{\widetilde{\alpha }}^{\beta }_{-m}]
{|B\rangle}^{({\rm osc})}\ =0,
\eea
\bea
({\alpha }^{i}_m-{\widetilde{\alpha }}^{i}_{-m}){|B\rangle}^{({\rm osc})}\ =0,
\eea
where the set $\{y^i\}$ indicates the position of the brane, and
$|B \rangle={|B\rangle}^{({\rm osc})}\otimes{|B\rangle}^{(0)}$.
According to the eigenvalues in Eq. (4) we deduce
the relation
\bea
p^\alpha =-\frac{1}{2\alpha'}[ ( \eta +4\omega
)^{-1}U ]^\alpha_{\;\;\beta} x^\beta .
\eea
This implies that along the worldvolume of the brane
momentum of the closed string depends on its center of mass
position. Therefore, in the presence of the tachyon field
the emitted closed string feels an exotic potential which affects its
evolution.

The solutions of the boundary state equations can be found
by the coherent state method. Thus, for the oscillating modes
we have
\bea
{|B\rangle}^{({\rm osc})}\ =\prod^{\infty }_{n=1}
{[\det Q_{(n)}]^{-1}}{\exp \left[-\sum^{\infty }_{m=1}
{\frac{1}{m}({\alpha }^{\mu }_{-m}S_{(m)\mu \nu }
{\widetilde{\alpha }}^{\nu }_{-m})}\right]\ }
{|0\rangle}_{\alpha }\otimes{|0\rangle}_{\widetilde{\alpha }}\;,
\eea
where the matrices are defined by
\bea
&~& Q_{(n){\alpha \beta }} = {\eta }_{\alpha \beta }+4{\omega
}_{\alpha \beta }-{{\mathcal F}}_{{\mathbf \alpha }{\mathbf \beta
}}+\frac{i}{2n}U_{\alpha \beta },
\nonumber\\
&~& S_{(m)\mu\nu}=(\Delta_{(m)\alpha \beta}\; ,\; -{\delta}_{ij}),
\nonumber\\
&~& \Delta_{(m)\alpha \beta} = (Q_{(m)}^{-1}N_{(m)})_{\alpha \beta},
\nonumber\\
&~& N_{(m){\alpha \beta }} = {\eta }_{\alpha \beta }
+4{\omega }_{\alpha \beta }
+{{\mathcal F}}_{{\mathbf \alpha }{\mathbf \beta }}
-\frac{i}{2m}U_{\alpha \beta }.
\eea
The advent of the normalization factor
$\prod^{\infty }_{n=1}{{[\det Q_{(n){\alpha \beta }}]}^{-1}}$
is anticipated by the disk partition function.
The boundary state associated with the zero modes possesses the
following feature
\bea
{{\rm |}B\rangle}^{\left(0\right)}
&=& \int^{\infty }_{{\rm -}\infty }
{\prod_{\alpha }{dp^{\alpha }}}\exp\left[i{\alpha }^{{\rm '}}\left(
\sum^p_{\alpha=0}{\left(U^{{\rm -}{\rm 1}}{\mathbf A}\right)}_{\alpha \alpha }
{\left(p^{\alpha }\right)}^{{\rm 2}}{\rm +}
\sum^{p}_{\alpha ,\beta {\rm =0},\alpha \ne \beta}{{\left(U^{{\rm -}{\rm
1}}{\mathbf A}+{\mathbf A}^T U^{-1}\right)}_{\alpha \beta }
p^{\alpha }p^{\beta }}\right)\right]{\rm \ \ }
\nonumber\\
&\times& \prod_i{\delta {\rm (}x^i}{\rm -}y^i{\rm )}
{\rm |}p^i{\rm =0}\rangle \otimes \prod_\alpha {\rm |}p^{\alpha }\rangle ,
\eea
where ${\mathbf A}_{\alpha \beta}=\eta_{\alpha \beta}
+ 4\omega_{\alpha \beta}$. The variety of the parameters in Eqs. (9)
and (11) enables us to adjust each part of the boundary state to a 
desirable state.

The total boundary state is
\bea
{|B\rangle}^{(\rm tot)}=\frac{T_p}{2}{|B\rangle}^{(\rm osc)}\otimes
{|B\rangle}^{(0)}\otimes{|B\rangle}^{(\rm gh)} ,
\eea
where ${|B\rangle}^{(\rm gh)}$ is the
boundary state corresponding to the conformal ghost fields
\bea
{|B\rangle}^{(\rm gh)}
=\exp\left[\sum^{\infty }_{m=1}{\left(c_{-m}{\widetilde{b\ }}_{-m}-b_{-m}
{\tilde{c}}_{-m}\right)}\right]\frac{c_0+{\tilde{c}}_0}{2}
\ |q=1\rangle\ \otimes|\tilde{q}=1\rangle.
\eea
In this state the ghost vacuum has been chosen in the picture
$\left(-1,-1\right)$.

Equations (9) and (12) define an effective tension for the brane, i.e.,
\bea
{\cal{T}}_p (T_p\;,\;{\cal{F}}\;,\;U\;,\;\omega)=
T_p \prod^{\infty }_{n=1}{[\det Q_{(n)}]^{-1}} .
\eea
\subsection{Modification of the boundary state}

It should be noted that the presence of the tachyon field in
the boundary state leads to an off-shell theory.
Therefore, we apply the procedure
of Ref. \cite{27} to deform it under the $PSL(2,R)$
transformation. Let $z$ be the complex coordinate of the points
within a unit disk which is the worldsheet of an emitted closed string
by the brane. Thus, the action of the $PSL(2,R)$ on the
complex coordinate $z$, is the mapping
$z \rightarrow w(z)=(az+b)/(b^{*}z+a^{*})$
where the complex variables $a$ and $b$ satisfy the relation
$|a|^{2}-|b|^{2}=1$. Since $PSL(2,R)$ is a subgroup of the
full conformal group of the disk, one can check
that this transformation preserves the shape and location
of the boundary of the disk.

Now we study the effect of this transformation on the closed
string oscillators which are defined by the following
contour integrals
\bea
\alpha^\mu_m &=& \sqrt{\frac{2}{\alpha'}}\oint_{C_z}
\frac{dz}{2\pi}z^m \partial_z X^\mu (z)
\nonumber\\
&=& \sqrt{\frac{2}{\alpha'}}\oint_{C_z}
\frac{dz}{2\pi}z^m \partial_w X^\mu (w)\frac{dw}{dz}.
\eea
The contour $C_z$ is a unit circle, i.e. the boundary
of the unit worldsheet disk of the emitted closed string.
There exists a similar formula for the left-moving
oscillators ${\tilde \alpha}^\mu_m$.
A mode expansion for $X^\mu $ also exists in terms of $w$
and ${\bar w}$ with the coefficients ${\alpha'}^\mu_m $
and ${\tilde {\alpha'}}^\mu_m$ exactly in the
feature of the first line of Eq. (15). Comparing the
corresponding coefficients we receive the relations
\bea
\alpha^\mu_0 &=& {\alpha'}^\mu_0 ,
\nonumber\\
{\tilde \alpha}^\mu_0 &=& {\tilde {\alpha'}}^\mu_0 ,
\nonumber\\
\alpha^\mu_m &=& M_{mn}(a, b){\alpha'}^\mu_n\;,
\nonumber\\
{\tilde \alpha}^\mu_m &=& - M^*_{mn}(a, b)
{\tilde {\alpha'}}^\mu_n\;,
\;\;\;\;\;\;m\;,\;n \in Z-\{0\} ,
\eea
where the matrix $M_{mn}(a, b)$ is defined by
\bea
M_{mn}(a , b)=\oint_{C_z} \frac{dz}{2\pi i} z^m
\frac{(b^* z+a^* )^{n-1}}{(az+b)^{n+1}}.
\eea
One can check that a creation (annihilation) oscillator
is expressed in terms of the creation (annihilation) oscillators.

In addition to the matter fields, the ghosts fields are also influenced  
by the $PSL(2,R)$ transformation. The procedure of Eqs. (15)
and (16) defines the following transformations for the oscillators 
of the conformal ghosts 
\bea
&~& b_m = N_{mn} b'_n ,
\nonumber\\
&~& c_m = P_{mn} c'_n ,
\nonumber\\
&~& {\tilde b_m} = - N^*_{mn} {\tilde b'_n} ,
\nonumber\\
&~& {\tilde c_m} = - P^*_{mn} {\tilde c'_n},
\eea
where the matrices are defined by
\bea
&~& N_{mn}= \oint_{C_z} \frac{dz}{2\pi i} z^{m+1}
\frac{(b^* z+a^* )^{n-2}}{(az+b)^{n+2}}\;,
\nonumber\\
&~& P_{mn}= \oint_{C_z} \frac{dz}{2\pi i} z^{m-2}
\frac{(b^* z+a^* )^{n+1}}{(az+b)^{n-1}}\;.
\eea

An appropriate redefinition of the boundary state,
consistent with the above transformations, is given by
\bea
&~& |{\bar B}\rangle = \int d^{2}a\;d^{2}b \;\delta (|a|^2-|b|^2-1)
|B_{a,b}\rangle^{\rm (mat)}\otimes |B_{a,b}\rangle^{\rm (gh)} ,
\nonumber\\
&~& |B_{a,b}\rangle^{\rm (mat)}=\frac{T_{p}}{2}\prod_{m=1}^{\infty}
[\det Q_{(m)}]^{-1}\exp \bigg{[}
\sum_{n=1}^\infty\sum _{j=1}^{\infty }
\sum _{k=1}^{\infty }\left(-\frac{1}{n}
\alpha^{\mu}_{-k}(S_{(n)jk}(a , b))_{\mu\nu}
\tilde{\alpha}^{\nu}_{-j}\right)\bigg{]}
\nonumber\\
&~& \;\;\;\;\;\;\;\;\;\;\;\;\;\;\;\;\times |0 \rangle_\alpha \otimes
|0 \rangle_{\tilde \alpha}\otimes |B\rangle^{(0)},
\nonumber\\
&~& |B_{a,b}\rangle^{\rm (gh)} = \left( |a|^2 + |b|^2\right)
\exp \left[\sum^\infty_{n=1}\sum^\infty_{j=1}\sum^\infty_{k=1}
\left(\Gamma_{(n)jk}(a,b)c_{-j}{\tilde b}_{-k}
- \Gamma^*_{(n)jk}(a,b)b_{-j}{\tilde c}_{-k}\right)\right]
\nonumber\\
&~& \;\;\;\;\;\;\;\;\;\;\;\;\;\;\;\times\frac{c_0+{\tilde{c}}_0}{2}
\ |q=1\rangle\ \otimes|\tilde{q}=1\rangle,
\eea
where for simplicity we removed the primes of the new oscillators.
The phrases $(S_{(n)jk}(a , b))_{\mu\nu}$ and 
$\Gamma_{(n)jk}(a , b)$ are given by
\bea
&~& \left(S_{(n)jk}(a , b)\right)_{\mu\nu}=-M_{-n,-k}(a , b)
S_{(n)\mu\nu}M^* _{-n,-j}(a , b),
\nonumber\\
&~& \Gamma_{(n)jk}(a , b) = - P_{-n,-j}(a,b)N^*_{-n,-k}(a,b).
\eea
The boundary state (20) will be used for calculation of the branes
interaction.
\section{INTERACTION OF TWO D-BRANES}

The interaction of two D-branes can be calculated in two
different but equivalent approaches: open string one-loop and
closed string tree-level diagrams. In the first approach which is a
quantum process, an open string is stretched between two D-branes.
It forms a cylinder via a loop diagram. In the second approach
which is
a classical process, a closed string is created by one D-brane. It
propagates in the transverse space between the two D-branes, and
then the other D-brane absorbs it. Here we use the second
approach for finding the interaction amplitude. Therefore, 
we calculate the overlap of the two boundary states via
the closed string propagator
\bea
{\cal{A}}=\langle B_1\left|D\right|B_2\rangle,
\nonumber
\eea
where $|B_1\rangle$ and $|B_2\rangle$ are total
boundary states corresponding to the D-branes and $``D "$ is
the closed string propagator 
\bea
D=2{\alpha }^{{\rm '}}\int^{\infty }_0{dt\ e^{-tH_{\rm closed}}}.
\eea
The closed string Hamiltonian is given by
\bea
H_{\rm closed}=H_{\rm ghost}+{\alpha' } p^{{\rm 2}}{\rm +\ 2}\sum^{\infty
}_{n{\rm =1}} {{\rm (}{\alpha }_{{\rm -}n}\cdot {\alpha }_n}{\rm +}
{\widetilde{\alpha }}_{{\rm -}n}\cdot {\widetilde{\alpha }}_n{\rm
)-}\frac{d-2}{6}.
\eea

The modified boundary state (20) can be applied for
obtaining the interaction amplitude of the
branes. After a long calculation, for parallel D$p$-branes we receive
the following amplitude 
\bea
{{\mathcal A}}&=&\frac{T^2_p \alpha' V_{p+1}}{4(2\pi)^{d-p-1}}
\prod^{\infty}_{n=1}{\det{\left[Q^\dagger_{\left(n\right)1}
Q_{\left(n\right)2}\right]}^{-1}}
\nonumber\\
&\times &\int d^2 a\;d^2b \;d^2a^{'}\;d^2b^{'}\delta
(|a|^2-|b|^2-1)\delta (|a'|^2-|b'|^2-1)
\nonumber\\
&\times & \left( |a|^2 + |b|^2\right)\left( |a'|^2 + |b'|^2\right)
\int^{\infty }_0 dt \bigg{\{} e^{(d-2)t/6}
\nonumber\\
&\times & \prod^\infty_{n}\prod^{\infty}_{j}\prod^{\infty}_{k}
\prod^{\infty}_{j'} \prod^{\infty}_{k'}\bigg{[}
\left(1-\Gamma^*_{(n)jk}(a,b)\Gamma_{(n)j'k'}(a',b')e^{-4nt}\right)^2
\nonumber\\
&\times &\left(\det (1-S^{(1)\dagger}_{(n)jk}(a , b)
S^{(2)}_{(n)j'k'}(a' , b') e^{-4nt})\right)^{-1}\bigg{]}
\nonumber\\
&\times& \frac{1}{\sqrt {\det  ( R^\dagger_1 R_2)}}
\left(\sqrt{\frac{\pi }
{\alpha' t}}\right)^{d-p-1} \exp \left(-\frac{1}{4\alpha't}
\sum_{i=p+1}^{d-1} \left(y^i_2-y^i_1\right)^2\right)\bigg{\}},
\eea
where $V_{p+1}$ is the common worldvolume of the branes, and
the symmetric matrices $R_1$ and $R_2$ possess nonzero elements only
along the branes worldvolume
\bea
&~& (R_l)_{\alpha \beta }=2\alpha'(-i{\mathcal M}_l
-iU^{-1}_l{\mathbf A}_l -i{\mathbf A}^T_l U^{-1}_l
+t{\mathbf{1}})_{\alpha \beta },\;\;l=1,2,
\nonumber\\
&~& {\mathcal M}_l{\mathbf =}\left( \begin{array}{ccc}
{\left(U^{-1}_l{\mathbf A}_l\right)}_{00} & \cdots  & {\mathbf 0}\\
\vdots  & \ddots  & \vdots  \\
{\mathbf 0} & \cdots  & {\left(U^{-1}_l{\mathbf A}_l\right)}_{pp}
\end{array} \right),
\nonumber\\
&~& ({\mathbf A}_l)_{\alpha \beta }{\mathbf =}\eta_{\alpha \beta }
+4({\mathbf \omega }_l)_{\alpha \beta }.
\eea
This amplitude includes contribution of all possible on-shell and
off-shell closed string states that the D$p$-branes can emit them.
The constant overall factor of the amplitude, behind the integrals,
indicates a hue of the strength of the interaction, which depends on the
parameters of the theory. The second determinant
is the contribution of the matter part oscillators, the factor 
including $\Gamma^* \Gamma$ comes from the ghosts and the other 
factors in the time integral are related to the zero modes of 
the string coordinates.
\subsection{The long-range force}

For distant D-branes only the massless states of closed string,
i.e. gravitation, dilaton and Kalb-Ramond states, exhibit a
considerable contribution to the interaction amplitude.
In other words, after a long enough time these massless 
states become dominant. Therefore,
we shall concentrate on the long-range force.

In the $26$-dimensional spacetime we acquire the following limit
\bea
&~& {\mathop{\lim }_{t\to \infty} e^{4t}} \prod^\infty_{n}
\prod^{\infty}_{j}\prod^{\infty}_{k}
\prod^{\infty}_{j'} \prod^{\infty}_{k'}\bigg{[}
\left(1-\Gamma^*_{(n)jk}(a,b)\Gamma_{(n)j'k'}(a',b')e^{-4nt}\right)^2
\nonumber\\
&~& \times \left(\det (1-S^{(1)\dagger}_{(n)jk}(a , b)
S^{(2)}_{(n)j'k'}(a' , b') e^{-4nt})\right)^{-1}\bigg{]}
\nonumber\\
&~& ={\mathop{\lim }_{t\to \infty} e^{4t}}+\sum^\infty_{j}
\sum^\infty_{k}\sum^\infty_{j'}\sum^\infty_{k'}
\bigg{[} {\rm Tr}\left( S^{(1)\dagger}_{(n=1)jk}(a , b)
S^{(2)}_{(n=1)j'k'}(a' , b')\right) 
\nonumber\\
&~& -2 \;\Gamma^*_{(n=1)jk}(a,b)\Gamma_{(n=1)j'k'}(a',b')\bigg{]}.
\eea
The application of this limit in the interaction amplitude
specifies the long-range force amplitude
\bea
{{\mathcal A}}_{\rm (massless)} &=& \frac{{T^2_p}
{\alpha'}}{4(2\pi)^{25-p}}
\prod^{\infty }_{n=1}{\det{\left[Q^\dagger_{\left(n\right)1}
Q_{\left(n\right)2}\right]}}^{-1}
\nonumber\\
& \times & \int d^2a\;d^2b\;d^2 a'\;d^2 b'\;
\delta (|a|^2-|b|^2-1)\delta(|a'|^2-|b'|^2-1)
\nonumber\\
& \times & \left( |a|^2 + |b|^2\right)\left( |a'|^2 + |b'|^2\right)
\int^{\infty }_0 dt \bigg{\{}
\nonumber\\
& \times &\bigg{[} {\mathop{\lim }_{t\to \infty } e^{4t}\
}+\sum^\infty_{j}\sum^\infty_{k}\sum^\infty_{j'}\sum^\infty_{k'}
\bigg{(}{\rm Tr}\left( S^{(1)\dagger}_{(n=1)jk}(a , b)
S^{(2)}_{(n=1)j'k'}(a' , b')\right) 
\nonumber\\
&~& -2 \;\Gamma^*_{(n=1)jk}(a,b)\Gamma_{(n=1)j'k'}(a',b')
\bigg{)}\bigg{]}
\nonumber\\
& \times & \frac{1}{\sqrt {\det  ( R^\dagger_1 R_2)}}
\left( \sqrt{\frac{\pi }{\alpha't}}\right)^{25-p}
\exp\left(-\frac{1}{4\alpha' t}
\sum_i \left(y^i_2-y^i_1\right)^2\right)\bigg{\}}.
\eea
Since the tachyon state has negative mass squared
the divergent term is corresponding to the exchange of the 
closed string tachyon, while the other terms represent the contribution
of the gravitation, dilaton and Kalb-Ramond states in the
interaction. Note that closed string emission
is independent of the locations of the branes, hence
the position factors do not change.
\section{INSTABILITY OF A ROTATING-MOVING D-BRANE}

D-branes in the presence of the tachyonic mode of the string
spectrum are under the experience of instability \cite{32}. This
is due to the rolling of the tachyon from an unstable maximum to
an incorrect vacuum (IR fixed point). This implies that adding the 
tachyon as a perturbation (deformation) to the theory leads to 
instability. In other words, tachyon condensation can make lower
dimensional unstable branes as intermediate states.

From technical point of view for constructing the tachyon
condensation we should take at least one of the tachyon's
elements to infinity. In the boundary states (9) and (11)
[or equivalently Eq. (20)] there
are four matrices $Q_{(n)}$, $\Delta_{(m)}$, $L=U^{-1}{\mathbf
A}+{\mathbf A}^T U^{-1}$ and the diagonal matrix 
$(U^{-1}{\mathbf A})_{\alpha \alpha}$, in which the tachyon 
has been inserted. 
So it is sufficient to take the limit of these matrices to evaluate
the behavior of the brane under the experience of tachyon
condensation.

Now we apply the limit $U_{pp} \to \infty$. According to the following
limit
\bea
{\mathop{\lim }_{U_{pp}\to \infty }
(U^{-1})_{p\alpha }=\mathop{{\rm \lim}}_{U_{pp}\to \infty }
(U^{-1})_{\alpha p}=0\ }, \;\;\;\alpha = 0,1, \cdot\cdot\cdot, p ,
\eea
the elements of the diagonal matrix
$(U^{-1}{\mathbf A})_{\alpha \alpha}
=(U^{-1})_{\alpha \gamma}{\mathbf A}^\gamma \; _{\alpha}
=(U^{-1})_{\alpha \gamma'}{\mathbf A}^{\gamma'}\; _{\alpha}
+(U^{-1})_{\alpha p}{\mathbf A}^p \;_{\alpha}$ with $\gamma' \neq p$
reduce to $(U^{-1})_{\alpha \gamma'}{\mathbf A}^{\gamma'} \;_{\alpha}$.
For $\alpha =p$ this element also vanishes and hence
we receive 
\bea
\sum^p_{\alpha =0}(U^{-1}{\mathbf A})_{\alpha \alpha}(p^\alpha)^2
= \sum^{p-1}_{\alpha' =0}
(U^{-1}{\mathbf A})_{\alpha' \alpha'}(p^{\alpha'})^2.
\nonumber
\eea
In the same way, the variables  
$\{L_{\alpha \beta}|\alpha , \beta =0,1,\cdot\cdot\cdot,p\;;
{\alpha \neq \beta}\}$ reduce to 
$\{L_{\alpha' \beta'}|\alpha' , \beta' =0,1,\cdot\cdot\cdot,p-1\;;
{\alpha' \neq \beta'}\}$.

The effect of the tachyon condensation on the factor 
$\prod^\infty_{n=1}[\det Q_{(n)}]^{-1}$
is given by the limit
\bea
&~& {{\mathop{\lim }_{U_{pp}\to \infty }\prod^{\infty}_{n=1}
{{\bigg [}\det {\bigg (}{\eta + 4\omega -{\cal{F}} +{\frac{iU}{2n}{\bigg )}}
_{(p+1)\times (p+1)}{\bigg ]}}^{-1}\ }\ }}
\nonumber\\
&~& =\prod^{\infty}_{n=1}{\ \ \frac{2n}{iU_{pp}}\left[
\det{\left( \eta + 4\omega -{\cal{F}} +\frac{iU}{2n}\right) }_{p\times p}
\ \ \right]^{-1}}.
\eea
The $p \times p$ matrix in the right-hand side is equal to the
$(p+1) \times (p+1)$ matrix in the left-hand side without
its last row and last column.

Now look at the matrix $\Delta_{(m)}$. In the limit $U_{pp} \to \infty$
its last row vanishes except $\Delta_{(m)pp}$ which goes to
$-1$. However, the elements of its last column, i.e.
$(\Delta_{(m)})_{\alpha' p}|_{\alpha' \neq p}$ do not vanish.
In fact, the resultant matrix possesses an eigenvalue ``$-1$''.
Therefore, in the diagonal form, it 
elucidates that the direction $x^p$ has been omitted from
the Neumann directions and has been added to the 
Dirichlet directions.

Adding all these together we observe that under the tachyon condensation
the D$p$-brane loses its $x^p$-direction and reduces to a
D$(p-1)$-brane. This implies that the rotation and motion 
of the brane do not induce a resistance against the instability, 
and hence collapsing of the brane takes place.

{\bf An example}

Now for our system consider the following special case, 
i.e. a rotating-moving D2-brane. Thus, under the limit 
$U_{22} \to \infty$ we obtain
\bea
{\mathop{\lim }_{U_{pp}\to \infty }}
\Delta_{(m)}=\left( \begin{array}{ccc}
\Delta_{(m)00} &
\Delta_{(m)01} &
\Delta_{(m)02}\\
\Delta_{(m)10} &
\Delta_{(m)11} &
\Delta_{(m)12}\\
0 & 0  & -1
\end{array} \right).
\eea
The eigenvalues of this matrix are as in the following
\bea
&~& \lambda_\pm =\frac{1}{2}\left(
\Delta_{(m)00}+\Delta_{(m)11} \pm 
\sqrt{(\Delta_{(m)00}-\Delta_{(m)11})^2 
+ 4\Delta_{(m)01} \Delta_{(m)10}}\;\right),
\nonumber\\
&~& \lambda_0 =-1.
\eea
According to the eigenvalue $-1$, after the tachyon condensation 
we obtain a new Dirichlet direction, i.e. $x^2$, while $x^0$ and $x^1$
remain Neumann directions. Similarly, the determinant of the matrix 
$\left(Q_{(n)}\right)_{3\times 3}$ reduces to
\bea 
{\mathop{\lim }_{U_{22}\to \infty }}\det Q_{(n)}
=\frac{i}{2n}U_{22}\det
\left( \begin{array}{cc}
Q_{(n)00} & Q_{(n)01} \\
Q_{(n)10} & Q_{(n)11}
\end{array} \right).  
\eea
The reduction also takes place for other tachyon-dependent
variables. Therefore, the D2-brane reduces to a
D-string along the $x^1$-direction. 
\section{CONCLUSIONS}

In this article we studied the boundary state of a closed
string, emitted (absorbed) by a rotating-moving D$p$-brane, in
the presence of the Kalb-Ramond field, a $U(1)$ gauge potential
and a tachyon field. 
The boundary state equations reveal that in the worldvolume
subspace the closed string momentum depends on its center
of mass position. This fact demonstrates that the tachyon field
exhibits a potential which acts on the closed string.

We deformed the boundary state by the $PSL(2,R)$ group.
The modified boundary state is a useful tool when the
tachyon field is presented in the system, and it is applicable for
both on-shell and off-shell closed strings. 

We obtained the interaction amplitude of two parallel 
rotating-moving D$p$-branes. 
The variety of the adjustable parameters controls
the treatment of the interaction. From the interaction amplitude
the long-range force was also extracted.
The constant overall factor of the interaction
amplitude indicates the strength of the interaction.
For mediating all closed string states or
only the massless states, which specify the texture of
the interaction, we received two different
strengths, as expected. For demonstrating the 
importance of branes interaction, for example, we can say that branes
interaction in the braneworld has been proposed as the origin
of inflation \cite{42}-\cite{44}, and an epoch in the early universe
that brings into begin the radiation-dominated big-bang.

Finally, we studied the tachyon condensation on a 
rotating-moving D$p$-brane via its
corresponding boundary state. We
observed that rotation and motion of the brane cannot 
prevent it from instability. Thus, the tachyon condensation
is terminated by the collapsing of the brane through 
reducing the brane dimension.

\end{document}